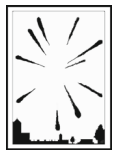

# Report

# Meteorite petrology versus genetics: Toward a unified binominal classification

Emmanuel JACQUET *

Institut de Minéralogie, de Physique des Matériaux et de Cosmochimie (IMPMC), Muséum national d'Histoire naturelle,
Sorbonne Université, CNRS, CP52, 57 rue Cuvier, 75005 Paris, France
*Corresponding author. E-mail: emmanuel.jacquet@mnhn.fr

**Abstract**–The current meteorite taxonomy, a result of two centuries of meteorite research and tradition, entangles textural and genetic terms in a less than consistent fashion, with some taxa (like "shergottites") representing varied lithologies from a single putative parent body while others (like "pallasites") subsume texturally similar objects of multifarious solar system origins. The familiar concept of "group" as representative of one primary parent body is also difficult to define empirically. It is proposed that the classification becomes explicitly binominal throughout the meteorite spectrum, with *classes* referring to petrographically defined primary rock types, whereas *groups* retain a genetic meaning, but no longer tied to any assumption on the number of represented parent bodies. The classification of a meteorite would thus involve both a class and a group, in a two-dimensional fashion analogous to the way Van Schmus and Wood decoupled primary and secondary properties in chondrites. Since groups would not substantially differ, at first, from those in current use de facto, the taxonomic treatment of "normal" meteorites, whose class would bring no new information, would hardly change. Yet classes combined with high- or low-level groups would provide a standardized grid to characterize petrographically and/or isotopically unusual or anomalous meteorites—which make up the majority of represented meteorite parent bodies—for example, in relation to the carbonaceous/noncarbonaceous dichotomy. In the longer term, the mergers of genetically related groups, a more systematic treatment of lithology mixtures, and the chondrite/achondrite transition can further simplify the nomenclature.

## INTRODUCTION

Any scientific discipline requires a conventional language to regulate the interaction of researchers with as little ambiguity as possible. Hence, the importance of a consensual taxonomy in meteoritics, which nowadays is de facto entrusted to the Nomenclature Committee of the Meteoritical Society (hereafter "NomCom"). Indeed, the approval by NomCom of the classification of a newly declared meteorite and inclusion in the Meteoritical Bulletin database (hereafter "MetBullDB") is also, indirectly, a judgment on the validity in abstracto of the taxon (I use here the word "taxon" (plural: taxa) to designate any set of meteorites in a classification, whatever its status in that classification (subgroup, group, clan, class, superclan, etc.)) it is assigned to.

Most classification terms of the current meteorite taxonomy were coined in the 19th century, as integrated in the Rose–Tschermak–Brezina taxonomy (e.g., Cohen, 1905; Mason, 1962; Tschermak, 1885) subsequently simplified by Prior (1920). Perforce, these terms had a foremost textural meaning. Still, at a time when a single parent body could be entertained for meteorites, be it the "missing planet" between Mars and Jupiter or a disintegrated terrestrial satellite (e.g., Boisse, 1847; Daubrée, 1867; Meunier, 1894, 1895), the taxa could have a genetic connotation as well, by relating to some specific layer in that parent body. By the 1960s, however, it had become clear that meteorites had multiple primary parent bodies (e.g., Fish et al., 1960; Lovering et al., 1957; Urey & Craig, 1953; Van Schmus & Wood, 1967) and that many Rose–Tschermak–Brezina taxa were solely textural. Thanks to progress in chemical and isotopic analyses, the taxonomy has since been refined in various chemical groups to match more the genetic relationships of meteorites (e.g., Weisberg et al., 2006).

With the surge of hot and cold desert finds in the last half century bringing new meteorite varieties to the attention of meteoriticists, the result is sometimes an inconsistent mixture of textural and genetic terms. Some Rose–Tschermak–Brezina taxa, more or less genetically interpreted, have swollen to encompass quite lithologically diverse objects, as for the shergottites (Walton et al., 2012), the angrites (Keil, 2012), or the diogenites (Beck & McSween, 2010). Conversely, other taxa, texturally interpreted, may comprise unrelated objects. For example, the Shallowater achondrite resembles aubrites, mineralogically speaking, and still appears as such in the MetBullDB; however, its mass-independent Cr isotopes and petrography are resolved from most aubrites (e.g., Keil, 2010; Zhu, Moynier, Schiller, Becker, et al., 2021) and it likely did not share their parent body. However, a reclassification as "achondrite, ungrouped" would erase from view that it is an enstatite achondrite, and it would be useful to allow a search of those specifically in databases, for example, in the hope of discovering new meteorite grouplets with this petrography. In fact, the "ungrouped" meteorites probably sample many more parent bodies than the established groups (e.g., Greenwood et al., 2020) and merit better taxonomic treatment, for example, as to isotopic affinities, in particular with the carbonaceous and noncarbonaceous (NC) superclans, which have acquired paramount importance in recent cosmochemical research (e.g., Kleine et al. [2020] and references therein).

This conflict between the textural and genetic aspects of the taxonomy certainly does not mean that one should be sacrificed to the other. In fact, it suggests that the taxonomy should cleanly separate these two hitherto intermingled aspects and become an explicitly binominal scheme, in the same way the two-dimensional Van Schmus and Wood (1967) classification decoupled primary and secondary properties in chondrites. This would allow us to ground classification on uniform principles throughout the meteorite spectrum. This is the subject of this paper. I will first expound the difficulties in defining the "group" concept as hitherto used in meteoritical literature, before introducing the guiding principles of a binominal (class-group) taxonomy which at last allows it a proper definition. It will be followed by a concrete proposition, mainly meant for illustration.

It is not the purpose of this work to be revolutionary, which would be the surest way to become a dead letter in so intrinsically conventional a matter as taxonomy. I have merely strived to slightly but methodically adjust (or set) the meaning of current terms, by way of systematically analyzing meteoritical taxonomic concepts. My touchstone has been to rationalize the system without materially changing most of the current literature practice at the present time, except to some extent for anomalous samples which would hereby gain a standardized taxonomic grid. In the long run, though, this could pave the way for a deeper (if gradual) simplification of the nomenclature, which is discussed before concluding.

## WHAT IS A GROUP?

In current usage, a meteorite "group" (or "chemical group") should ideally comprise meteorites derived from the same celestial body. However, this is difficult to ascertain from the ground. Cosmic ray exposure age statistics for a given taxon (e.g., Herzog & Caffee, 2014) can only indicate that some ejection age peaks mostly derive from a single primary parent body, but cannot decide for the taxon at large. For instance, the ~7 Ma peak for H chondrites essentially applies to H4 and H5 (Herzog & Caffee, 2014), so cosmogenic isotope data *alone* would statistically allow a separate origin for H3 and H6 (or parts thereof). The frequent occurrence of breccias' mixing clasts from different lithologies (e.g., genomict breccias for H chondrites) may support a common origin, but this may likewise hold only for a fraction of them. For any single meteorite, the association with a given group is based on textural, chemical, and/or isotopic similarities. This assumes a restricted range of variation inside the parent body, but also that separate parent bodies had resolved properties.

However, the solar protoplanetary disk where the planetesimals accreted probably presented a continuum of compositions (e.g., Jacquet et al., 2019; Pignatale et al., 2018) save perhaps for one or a few hiatuses such as the one separating the NC and C superclans (e.g., Desch et al., 2018; Fukai & Arakawa, 2021). So two distinct primary parent bodies which formed spatially and temporally close by in the disk may present close (isotopic, chemical, and/or textural) compositions. Now the present asteroid Main Belt, containing ~5 × 10$^{-4}$ Earth masses, must host but a very partial sampling of the planetesimals that accreted in its neighborhood, or the solar system at large, hence why, in general, meteorites form discrete ensembles resolved from one another in compositional space (Jacquet et al., 2016), numbering perhaps in the 95–148 range (Greenwood et al., 2020) (see, e.g., Fig. 1 for pallasites). The surviving parent bodies may present "chance" clustering in compositional space—forming the "clans" of Weisberg et al. (2006) (see Fig. 2). As such, "clans" may have little intrinsic meaning. While they sample some common space–time section of the protoplanetary disk—a "reservoir" in cosmochemical parlance—the bounds of that reservoir, if at all defined, would be arbitrary[1] and not necessarily have any astrophysical significance. For instance, the gap between ordinary (OCs) and enstatite chondrites (ECs) may well be bridged in reality by reduced chondrites hitherto only known as unique samples or grouplets (e.g., Néal & Lipschutz, 1981; Pourkhorsandi et al., 2017; Weisberg et al., 2015).

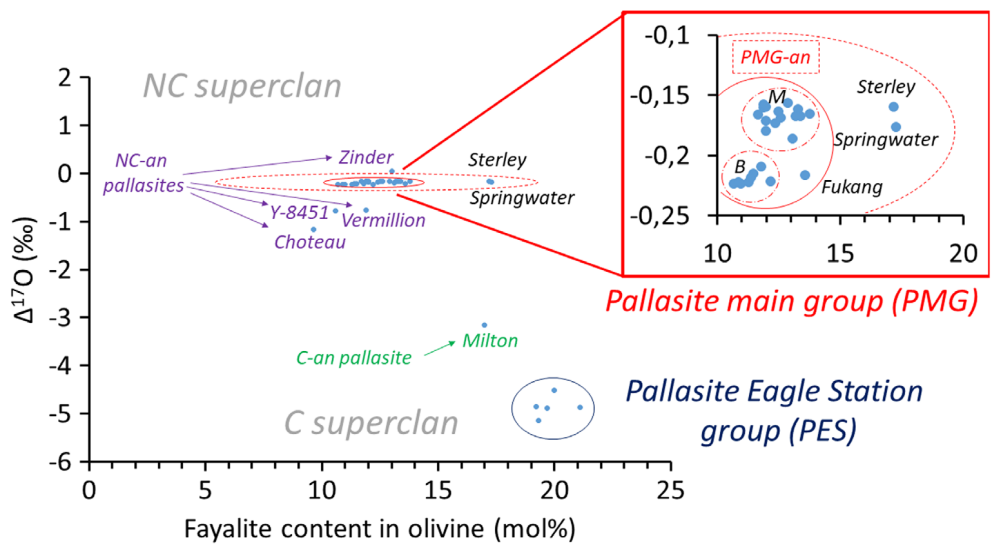

Fig. 1. Clustering in compositional (here isotopic-mineral chemical) space illustrated by pallasites. The pallasite main group (PMG) and Eagle station group (PES) are circled with thin solid lines. The blowout shows that the PMG may be actually bimodal (Ali et al., 2018), illustrating how relative the concept of group must be when empirically understood. With $\Delta^{17}O$ indistinguishable from the PMG, Springwater and Sterley belong to its "halo" (schematically bounded by a dotted line) and may be held to be PMG-an (even though the MetBullDB lists Sterley as simply PMG). Other pallasites can be tied neither to PMG or PES and are only broadly affiliated to the C or NC superclan (see the Anomalous Meteorites section for more on the notation of anomalies). Data from Ali et al. (2018); Wasson and Choi (2003); Clayton and Mayeda (1996); and MetBullDB. (Color figure can be viewed at wileyonlinelibrary.com.)

Now, the clustering in compositional space of several parent bodies may be so tight that the compositional variability of the derived meteorites resembles that of a single primary parent body, and distinguishing between the two possibilities becomes arduous for the meteoriticist. A case in point is the CV-CK clan. Wasson et al. (2013) proposed that CK chondrites derived from metamorphosed regions of the CV parent body. Yet recent chromium isotopic data indicate that they are isotopically resolved, suggesting different parent bodies of origin (e.g., Zhu, Moynier, Schiller, Alexander, et al., 2021), unless some heterogeneous accretion can be assumed. In fact, the CV group itself could be composite in parent body provenance, as Gattacceca et al. (2020) suggested, on the basis of chondrule size and matrix abundances, for the "reduced" and "oxidized" subgroups $CV_{red}$ and $CV_{ox}$ (itself a combination of the older $CV_{oxA}$ and $CV_{oxB}$, named after Allende and Bali, respectively; Weisberg et al., 1997) originally distinguished by their secondary mineralogy. So depending on the authors, the elusive "parent body taxon" of, say, Allende could be either a clan, a group, or a subgroup (or even a "sub-subgroup")!

Obviously, the exact *genetic* implications of a taxon are not something an empirical taxonomy should decide. Here and throughout, "genetic" relates to spatiotemporal proximity of formation in the early solar system (either within a single primary parent body or a more or less broad "reservoir" in the disk). How, then, can a group be *empirically* defined? Obviously, there is no such thing as an interbreeding criterion as for biological species.

For chondrites, Krot et al. (2014) provide a definition of a chemical group, similar to that given by Weisberg et al. (2006) and Rubin and Ma (2021) and which may be held to be consensual:

---

[1]The phrase "parental reservoir" can have a more definite meaning for *one* given parent body in designating the composition of the immediate neighborhood (in space and time) of its accretion, if rapid (homogeneous) accretion can be assumed, without any further assumption on the lengthscale of variation of the disk composition.

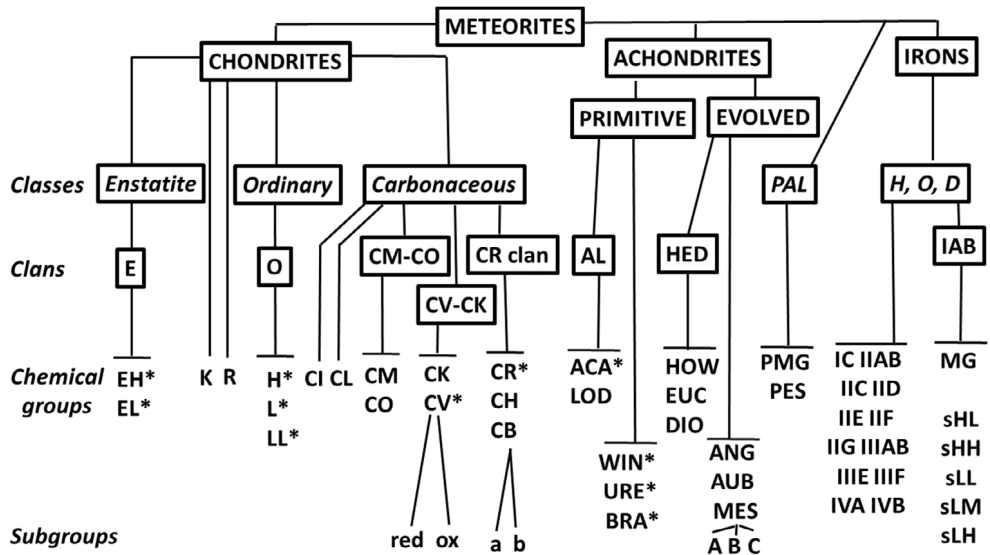

Fig. 2. The traditional hierarchical classification system for asteroidal meteorites. Only chemical groups recognized in MetBullDB are listed. Higher level taxa (classes and beyond) are petrological in character but chemical groups, in particular, have genetic significance, even though its nature cannot be exactly ascertained (e.g., number of primary parent bodies). Modified after Weisberg et al. (2006). ACA = acapulcoite, AL = Acapulco-Lodran clan, ANG = angrite, AUB = aubrite, BRA = brachinite, DIO = diogenite, HOW = howardite, LOD = lodranite, MES = mesosiderite, PAL = pallasite, PES = pallasite Eagle Station group, PMG = pallasite main group, URE = ureilite, WIN = winonaite. The asterisks indicate that the groups in question do not entirely fit in the higher level taxa indicated (e.g., some ureilites and brachinites may be cumulates and thus evolved achondrites; Day et al., 2012; Goodrich et al., 2009; many chondrite groups may also extend to primitive achondrites, sometimes subsumed under petrologic type 7). The iron classes (H = hexahedrite, O = octahedrite, D = ataxite) are mentioned for completeness but have fallen in disuse, and the iron chemical groups do not subdivide but straddle them. The IAB complex ("IAB" in the diagram; Wasson & Kallemeyn, 2002) is subdivided in a main group (MG) and several so-called subgroups whose suffixes ("sHH" for "IAB-sHH," etc.) are indicated there.

Conventionally, a chondrite "group" is defined as having a minimum of five unpaired chondrites of similar mineralogy, petrography, bulk isotopic properties, and bulk chemical compositions in major, nonvolatile elements.

One can only subscribe to the membership threshold as a safeguard against misunderstood outliers or possible extensions of preexisting groups. One may echo the recommendation of Scott and Wasson (1975), in their review of iron classification, that, ideally, a group should show internal trends distinct from intergroup trends (in the case of most irons discussed specifically by Scott and Wasson [1975]: fractional crystallization versus volatile element depletion).

Now, "similar" in the above definition does not mean "indistinguishable"; one only needs to think of the range in O isotope composition in groups such as CM chondrites, or the range in olivine compositions within the different ordinary chondrite groups. Rather "similar" refers to some restricted subset of compositional space. Then, absent quantitative similarity metrics, the definition above, taken literally, is not specific to what Krot et al. (2014) or Weisberg et al. (2006) called "groups." We could go over the previous example again, and see that CV-CK, CK, CV, $CV_{red}$, $CV_{ox}$, $CV_{oxA}$, $CV_{oxB}$ all satisfy this definition. Likewise, the ordinary chondrites, while of course showing more variability than their individual component chemical groups H, L, and LL, are still similar (though not identical) to one another, when compared to nonordinary chondrites, and represent a specific region of isotopic and chemical space not overlapped by the latter. Basically, subgroups, chemical groups, clans, even superclans (C and NC) are groups by the traditional definition. There is little hope to define a priori quantitative criteria, for example, maximum standard deviations of some select compositional parameters, that would be suitable for all cases. Which one should we choose? Obviously, oxygen isotopes, so useful for stones, would be hardly relevant for most irons.

Weisberg et al. (2006) noted that the genetic connotation of the word "group" was looser for differentiated meteorites. Howardites, eucrites, and diogenites are ranked as three different "groups" yet share the same parent body, so the parent body taxon is rather the HED clan. Rubin and Ma (2021) systematically use the word "clan" to mean putative parent body taxa for differentiated meteorites. So "groups" refer there to specific lithologies inside a parent body.

Conversely, other taxa seem to refer to specific lithologies from different origins. For example, "pallasite" must represent several parent bodies, both in the carbonaceous and NC reservoir, and thus is merely a "textural term" (Weisberg et al., 2006). Weisberg et al. (2006) also mention the case of the Shallowater "aubrite" cited in the introduction, to which we may add the eucrite-like meteorites such as Pasamonte and Ibitira, which are isotopically resolved from most HED (e.g., Mittlefehldt, 2015). Likewise, the somewhat variable isotopic signatures of brachinites (Keil, 2014) may point to several parent bodies.

Now a genetic meaning can be restored by refining the taxonomy. Pallasites are already subdivided in a pallasite main group and a pallasite Eagle Station group, not to mention ungrouped pallasites (e.g., the pyroxene pallasites of Boesenberg et al., 1995). Isotopic outliers in other groups could be excluded therefrom; Mittlefehldt et al. (2022) reserve the word "eucrite" to isotopically normal HED basalts/gabbros; likewise, the "aubrite main group" of Zhu, Moynier, Schiller, Becker, et al. (2021) could be appropriated the "aubrite" term. In fact, recent (re)classifications already prefer to use "achondrite, ungrouped" even in the presence of a clear textural affinity to known groups (e.g., the eucrite-like NWA 011 or the diogenite-like NWA 6693). Some current groups, such as brachinites, may be "too composite" for a main subdivision to be readily distinguished, but would nonetheless have genetic meaning (if beyond a single parent body scale) given the isotopic proximity of their members. So it may seem that the Rose–Tschermak–Brezina and like terms (at least those etymologically associated with a type meteorite) may be decidedly given a genetic sense—if perforce noncommittal as to the implied number of primary parent bodies.

This, however, would only hold at the lowest levels of the taxonomy if understood as a single hierarchical system. Indeed, in Fig. 2 (see also those of Krot et al., 2014; Weisberg et al., 2006), the highest level taxa are petrological in character (e.g., differentiated versus primitive meteorites). This does no justice to genetic affinities. For example, aubrites and ECs remain widely separated in the arborescence even though they may derive from the same E reservoir. Conversely, pallasites, although relatively low level in the hierarchy, have feet in both NC and C reservoirs. The arborescence itself is sometimes somewhat approximative: Ureilites should in principle be split between primitive (olivine–pigeonite and olivine–orthopyroxene) and evolved (olivine–augite) achondrites (Goodrich et al., 2004, 2009), and the same may hold for brachinites as well (Day et al., 2012). Likewise, some chondrite groups may extend to the primitive achondrite category depending on where the line is drawn (e.g., Tomkins et al., 2020) (Fig. 2).

Weisberg et al. (2006) sketched in their fig. 8 a "genetic" classification scheme as an alternative to the conventional (and texturally inspired) scheme illustrated in their fig. 1 but lost track of the petrological classification. Warren (2011b), who proposed to place the C/NC dichotomy at the highest level of the classification, retained the chondrite/achondrite distinction but had to duplicate it within both superclans. I suggest that we do not have to choose and can make the taxonomy explicitly binominal to make way for both logics.

## PRINCIPLES OF A BINOMINAL SCHEME

### Groups and Classes

In his "tabular classification," Prior (1920) systematically distinguished between "classes" and "groups," the former corresponding to varying metal/silicate ratio and the latter to varying redox levels (Fe/Ni ratio in metal; FeO/MgO ratio in silicate). The groups had genetic significance in the Prior (1916) picture of progressive oxidation anticorrelating with depth in a parent body. This conceptualization of the taxonomy was later adopted by Yavnel (1958), who ascribed his groups to different parent bodies. Mutatis mutandis (e.g., with isotopic ratios having superseded redox levels as genetic signatures), I propose to revive this two-dimensional logic by redefining "groups" and "classes" in a unified way across meteorite categories as follows:

A *group* is a genetically connoted taxon (hence the mnemonic    *G*roup = *G*enetics), whose empirical definition is only slightly modified from the letter of Krot et al. (2014) as: "a set of meteorites characterized by similar isotopic signatures and, possibly, similar petrologic characteristics (bulk chemistry, mineralogy, and/or texture)." A kinship in some isotopic ratios (at least two, to break the degeneracy between time and heliocentric distance) is thus required for the genetic aspect of a group,[2] but petrographic aspects may have various degrees of importance depending on its breadth (i.e., hierarchic level) and historical practice: for example, the definition of CV chondrites obviously contains petrographic criteria (e.g., chondritic texture, CAI abundance, etc.), but that of the set of all C meteorites, primitive or differentiated (definable, e.g., by $^{54}Cr$ and $^{50}Ti$ enrichments and/or a Mo isotopic signature; e.g., Kleine et al., 2020), does not. The word "similar" is understood to refer to some connected region of compositional/textural space (e.g., a "zero-dimensional" patch whose breadth is dominated by analytical uncertainties, some one-dimensional trend, etc.). This obviously comprises the hitherto recognized "chemical groups" of chondrites (e.g., "H") or irons (e.g., "IIAB"), as well as the genetically understood Rose–Tschermak–Brezina (and like) achondrite groups (e.g., "eucrites," if isotopic outliers are excluded; "winonaites," etc.).

---

[2]At least its *definition*, for in practice, *assignment* of a new meteorite to a preexisting group may be simply based on close mineral chemical correspondence.

A *class* is a primary petrographic taxon (i.e., referring to the rock prior to any secondary alteration). In principle, it should be determinable based on textures and mineralogy (optical or scanning electron microscopy) alone, but it may occasionally be somewhat interpretative (if secondary effects are significant). This word has been already used with the same effective denotation for chondrites (e.g., "enstatite chondrites"; Weisberg et al., 2006) and irons (precisely, their *structural* classes; e.g., "octahedrites"; Buchwald, 1975).

Thus, a meteorite classification would feature a class and a group, for example, "hexahedrite" and "IIAB," respectively, for the North Chile iron. There are thus a "classwise" classification and a "groupwise" classification. At the highest levels, the distinction between primitive and differentiated meteorites is a classwise distinction, that between NC and C meteorites is a groupwise distinction.

Since we no longer deal with a single hierarchical system, but a two-dimensional one, it should be emphasized that classes as understood here *are not* sets (nor unions) of groups, no more than Van Schmus and Wood (1967) petrologic types were. In fact, a group may straddle several classes. For example, the diogenite group comprises dunites, harzburgites, orthopyroxenites, and norites (e.g., Beck & McSween, 2010). This I call "class transversality." In practice, however, one class will generally dominate for any given group, and one might define it a priori as a default class (e.g., "orthopyroxenitic achondrites" for diogenites), with specific petrographic information (e.g., olivine mode in a diogenite) required if a declaration claims a different assignment, so as to maintain consistency between classifiers. If a meteorite belongs to the default class of its group, that class could be omitted in database classification entries (such as the MetBullDB).

The above definitions should also make clear that classes have no genetic significance and thus do not have to be confined to meteorites of close solar system provenance. For example, as mentioned earlier, the pallasite class comprises samples from both the C and the NC superclans. Moreover, although classes are petrographically defined, they are not meant to be so refined as to capture all textural properties of the member meteorites (no more than petrographic "type 2" means the same aqueous alteration effects in CR and CM chondrites). For example, among orthopyroxenitic achondrites, aubrites and diogenites are certainly quite distinct even from a purely mineralogical point of view. The touchstone of classes is rather that they capture the important nuances inside each group (e.g., for diogenites, the distinction among orthopyroxenitic, noritic, and dunitic varieties, as mentioned above).

The binominal scheme is primarily meant as a primary classification. Secondary effects such as aqueous alteration, thermal metamorphism, shocks, and weathering make further taxonomic dimensions already abstracted from groups in previous literature (Stöffler et al., 1991, 2018; Van Schmus & Wood, 1967; Wlotzka, 1993) and which will not generally be our focus here. Yet petrologic types may be understood to subdivide groups in taxa like "H5" or "H4" which themselves may be viewed as groups (since the petrographic character of groups under the present redefinition does not have to be primary, unlike for classes).[3] This technicality may be more meaningful than it seems because the genetic relationships between different petrographic types of a conventional chemical group must remain perforce somewhat conjectural. Van Schmus and Wood (1967) had from the outset denied simple genetic relationships between their C1, C2, C3 chondrites, which were to later break down in several chemical groups (e.g., Van Schmus & Hayes, 1972). Later, Sears et al. (1982) showed that their E4 and E5 (both recast as EHs) had a parent body different from their E6s (reclassified as ELs). More recently, had CK chondrites been recast as (more or less metamorphosed) CVs following Wasson et al. (2013), Zhu, Moynier, Schiller, Alexander, et al. (2021)'s conclusions would have amounted to divorcing the "CV4," "CV5," "CV6" from most CV3.

---

[3]This, in fact, equally applies to classes since any class–group intersection is, in intension, a group (with the class properties factoring into the petrologic criteria of the group). For example, the subset of IIAB irons which are hexahedrites is itself a group—and in fact formerly had a proper name, "IIA"—so one could be content with a monistic groupwise classification as "IIAB hexahedrite." The point of conceptually distinguishing classes is to give them a uniform terminology which avoids unnecessary inflation in nomenclature. Class–group intersections are sufficiently identified by derived names built after the intersecting taxa (e.g., "dunitic diogenites") and do not require proper names.

## Hierarchic Agnosticism

While the terms "class" and "group" are quite familiar, their meaning is widened here compared to recent literature because they are no longer tied to specific levels in some single hierarchy (as they were in the traditionally minded Fig. 2 where "chemical groups" lay between the "subgroup" and the "clan" levels, and "classes" just above). Rather, since the definitions of the previous subsection can apply to any level of precision of petrographic or isotopic properties, one can form a hierarchy of classes and a separate hierarchy of groups. That is, a class can be a subset of (i.e., be included in) another (higher level) class, and a group can be a subset of another group. Thus, for example, "irons" are viewed as a class, but "octahedrites" also make a (lower level) class, included in the former, in its own right. Likewise, a subgroup is a group, a chemical group is a group, a clan is a group, a superclan is a group. (Here and henceforth, the phrase "chemical group" denotes what past literature has called group [with or without the adjective], whenever the distinction will be useful.) Hence, H chondrites form a group, but so do ordinary chondrites as a whole (the "OC group"), or the even larger set of all NC meteorites (the "NC group").

However meaningful it is to judge whether a genetically minded taxon is sufficiently well defined to warrant systematic evaluation in MetBullDB (as recently with $CV_{red}$ and $CV_{ox}$; Gattacceca et al., 2021), there is no point (and no objective criterion) in granting official hierarchical distinctions to such taxa, and call some "chemical groups," other "subgroups," or "clans."[4] This is because, as argued in the previous section, we *as classifiers* have to remain agnostic as to the exact genetic significance (viz. number of represented parent bodies or size/lifetime of the parental reservoir in the disk) of our taxa. This in no way prevents us *as theorists* from endorsing or rejecting the single parent body provenance of one group or another in research articles. Certainly, a prevalent opinion in the cosmochemical community that some proposed taxon may represent a unique parent body will contribute to its formal recognition as a group, but this needs not be made an official requirement. It is worth remembering that Van Schmus and Wood (1967), who first systematically conceptualized the chemical groups for chondrites, had deemed them useful regardless of the truth of the implied simple genetic implications—which in fact we saw they already denied for the C chondrites (CCs).

At any rate, the level of a taxon in a hierarchy is a function of the number of recognized divisions above and/or below it. For example, ordinary, enstatite, and carbonaceous chondrites may appear on the same level in the hierarchy of Fig. 2, but this would change if one introduces the NC/C division immediately below the chondrite taxon: then, OCs and ECs become mere subdivisions of the NC chondrites while CCs would remain at a higher level. So the "level" cannot be granted intrinsic value except for "local" comparison of nested taxa (with e.g., H being lower level than OC, which is itself lower level than NC chondrites).

So when occasionally referring to "clans," "chemical groups," and "subgroups" herein, I will simply refer to what the literature has called such (see e.g., Fig. 2) but I consider these "grandfathered" terms as merely historical accidents that cannot be given any intrinsic empirical definition—beyond the historical requirements that a clan should contain several groups previously recognized by NomCom and a subgroup should be a (strict) subset of another group previously recognized by NomCom. The *concepts* of "clans," "chemical groups," and "subgroups" are hence not part of the binominal formalism, even though the *names* can be used for convenience. The name "superclan" will be reserved to the NC and C groups.

Although a meteorite can belong to several (nested) classes or groups, the MetBullDB should assign it to the most precise ones known, that is, in principle those at the lowest level of the taxonomy (so e.g., "CK chondrite" rather than simply "CC"). Exceptions include samples insufficiently characterized from the point of view of present taxonomy, although they may have been from that at the time of their declaration (e.g., many "CV3" classified before NomCom recognized the $CV_{ox}$/$CV_{red}$ distinction) and anomalous meteorites which can only be associated with high-level groups or classes (which are discussed in the next subsection). Yet in the majority of the cases, groups effectively displayed by the MetBullDB under a binominal scheme would be those hitherto recognized as such (i.e., "chemical groups") or subgroups.

## Anomalous Meteorites

As in previous practice for groups (e.g., Krot et al., 2014), a taxon may be formally granted a new, proper name only if it comprises at least five unpaired meteorites. This does not apply to *derived* names, that is, names built after preexisting taxon names, for example, for transitional or polymict samples discussed in the next subsection.[5] This, however, *does* apply to "subgroup" names such as $CV_{ox}$ or $CV_{red}$ since the added subscripts do not refer to preexisting taxa (reassuringly, these and all other "subgroups" mentioned in Fig. 2 do satisfy the threshold already). Proposed taxa not meeting the five-member threshold yet may be informally referred to as "grouplets" (e.g., the "G chondrites"; Ivanova et al., 2020; Weisberg et al., 2012, 2015) and "classlets" (e.g., andesites; Barrat et al., 2021). Officially though, their members cannot be associated with any recognized low-level class and/or group and must be considered "anomalous" (class- and/or groupwise).

---

[4]This is not unlike some calls for unranked taxonomy in biology (e.g., Mishler, 1999) doing away with the need to call some monophyletic taxa "genera," "orders," "classes," "phyla," etc., absent real definitions thereof, while many other equally monophyletic taxa (in intermediate position) have no such rank anyway.

[5]Another example of "derived" naming mentioned at the end of section Groups and Classes is that of the "chemical group and petrologic type" groups, for example, CL4 chondrites. Although only three CL4 are known (e.g., Metzler et al., 2021), the five-member rule is not violated because the name is built after the "CL" and "type 4" taxa which individually satisfy the threshold.

Now, as to classes, no meteorite will be conceivably so mineralogically anomalous as not to fit in at least the broad classes "chondrites," "achondrites" (or "stones" for ambiguous cases), or "irons." So they may be ranked as "anomalous chondrites," "anomalous achondrites," or "anomalous irons" (i.e., *structurally* anomalous irons). The Erg Chech 002 andesite (Barrat et al., 2021) would thus be an anomalous achondrite, for example. Lower levels of anomalies are possible: Ibitira would be reckoned among the basaltic achondrites because of its fine-grained pyroxene–plagioclase-dominated mineralogy, but would differ from its normal members in its abundance of vesicles (e.g., Mittlefehldt et al., 2022) and thus be an "anomalous basaltic achondrite." So the level of the anomaly would be measured by the level of precision of the associated "root" class ("basaltic achondrite" being here more precise than "achondrite," and thus Ibitira being in a sense "less anomalous," classwise, than Erg Chech 002).

The same anomaly hierarchy can be applied to groupwise classification. For example, Burnwell is a type 4 ordinary chondrite slightly more reduced ($Fa_{15.8}$) than normal H chondrites ($Fa_{16.5–20.8}$) and has been classified as a H4-an. This basically expresses the expectation that new meteorites may extend the H field and reveal Burnwell as a bona fide (normal) member of the group. NWA 13411, while still petrographically an ordinary chondrite, has an average $Fa_{8.35}$ far removed from H and a fortiori any other recognized ordinary chondrite group, and is thus ranked as O5-an.[6] Indeed, it is conceivable that someday a sufficient number of related chondrites will be found to create a new group within the OC clan, which would make NWA 13411 a normal member of the latter. That is, *being a normal member of some high-level group is the same as belonging to one of its established component groups*. Likewise, an anomalous type 2 CC such as NWA 5958 (Jacquet et al., 2016) should be ranked as C2-an. The current

MetBullDB (and Jacquet et al., 2016) calls it a "C2-ung," because the C superclan has not hitherto been considered as a group (which alone could be attributed an "anomalous" suffix). The binominal scheme, however, enfranchises "superclans" and "clans" as groups in their own right and thus makes any distinction unjustified and moot.[7] No distinction of the sort had been anyway made as late as the time of the fourth edition of *Catalogue of Meteorites* (Graham et al., 1985) which only knew of "anomalous" ("chanom," "acanom," "iranom"), not "ungrouped" meteorites.

Now there are arguably meteorites which are not even remotely associated with any group whatsoever, such as many chemically anomalous irons, so they could still be called "ungrouped" (as a distinction from classwise anomalies). Yet, increasingly systematically, anomalous meteorites are subjected to isotopic analyses (e.g., Spitzer et al., 2021), which may reveal affinity, if not to known "clans," at least to the NC or the C superclans. So, for instance, the eucrite-like NWA 011 meteorite, which plots in the C field in a $\delta^{50}Ti$-$\delta^{54}Cr$ diagram (e.g., Warren, 2011a, 2011b) should be classified as "basaltic achondrite, C-an", which would be more informative than its current classification as simply "achondrite, ungrouped." We have likewise "NC-an pallasites" and "C-an pallasites," as depicted in Fig. 1. Given the growing importance of the C/NC dichotomy in cosmochemical research, making the affiliation to either superclan visible or searchable in the MetBullDB whenever it is known for (genetically) anomalous meteorites would be certainly most advantageous. This is what the binominal scheme allows without any further conceptual addition.

## Lithology Mixing

Assuming a group- and classwise taxonomy for unbrecciated meteorites, breccias may be systematically classified in terms of their component lithologies without inflating the terminology. At a minimum, a "monomict" or "polymict" qualifier may be used (if, say, >90 vol% of clasts can be subsumed in the stated taxa). If several individually classified lithologies exceed 10 vol%, their taxa shall be concatenated with a hyphen ("–") in the group- and/or classwise classification. For example, howardites, which *are*, in theory, defined by diogenite and eucrite clasts both exceeding 10 vol% (e.g., McSween et al., 2011), could be explicitly recast as "diogenite-eucrite" (groupwise). The microbreccia Kaidun might be described groupwise as a C1-CM-E chondrite, if bulk O isotopic measurements adequately identify the dominant lithologies (Zolensky & Ivanov, 2003).

---

[6]The MetBullDB says "OC5-an," but the additional "C," redundant with the indication of a chondrite petrologic type, can be dropped in consistency with E and C chondrites (e.g., E3-an LEW 87223 or C2-an NWA 5958).

[7]So the classification of NWA 7891 as a "CV3-an" would not depend on the status of the $CV_{ox}$ and $CV_{red}$ subgroups. In current practice, if they are someday recognized as chemical groups, this "CV3-an" would have to become "CV3-ung" without any change in its intrinsic characterization (notwithstanding the possibility that it could be found to belong to the particular "halos" of either one and become a $CV_{ox}$3-an or a $CV_{red}$3-an).

The hyphen convention is inspired from that already in use for genomict breccias, where a "H3-5" notation compactly designates a breccia of H chondrite clasts spanning petrographic types between 3 and 5 (even though a "H3-H4-H5" or "H3-H5" notation would be more precise). It may be well to standardize the order, for example, alphabetically, to avoid variants in databases (e.g., "diogenite–eucrite" and "eucrite–diogenite"), unless it is meant to reflect the relative modal abundances (which however would be more liable to nonrepresentative sampling).

It is worth distinguishing between the hyphen ("–") and the often-used slash ("/") notations, as the latter symbol should be restricted for lithologies that are truly transitional (i.e., in unbrecciated state) between the taxa in question. For example, L/LL chondrites are those (anomalous) ordinary chondrites which have mineral composition averages in both L and LL fields (for different phases, e.g., metal and olivine) and/or some in-between (Rubin, 1990). So, a lunar "mingled" breccia should be an "anorthosite–basalt breccia" rather than "Lunar (bas/anor)" as in the current MetBullDB.

This convention can be extended to mixtures of (preexisting) lithologies which are not (postlithification) breccias: The shergottite EETA 79001, which features a contact between a basaltic and an olivine–phyric lithologies (e.g., McSween & McLennan, 2014), can be ranked classwise as a basalt-(olivine basalt). Silicate-bearing irons whose silicate inclusions may have been introduced into molten metal (e.g., Scott, 2020) could be characterized in a hyphenated way if metal and silicate lithologies are individually characterized class- and groupwise (e.g., Dey et al., 2019). Isheyevo, currently classified as CH/CB$_b$, which shows distinct layers (if with sometimes gradational boundaries) of CH- and CB$_b$-like materials (e.g., Ivanova et al., 2008) could be more precisely reclassified as a (CH-an)-(CB$_b$-an). The "hyphen" convention is thus agnostic on the nature of the relationship and the mixing between lithologies.

## A PROPOSED CLASSIFICATION

The principles above have been laid out in abstracto, if illustrated with scattered examples. For more concreteness, I will now sketch an actual classification which should be mainly viewed as a starting point, for illustrative purposes. Again, it is not my intent to be revolutionary. The lowest level groups, which meteoriticists chiefly attend to, will be exactly those in current use (with the understanding that isotopic outliers are excluded as discussed at the end of the section What Is a Group?). Classes which will serve to characterize unusual meteorites (as well as illuminate the first-order petrographic characteristics of established groups) will, however, have to be somewhat more innovative (if with self-explanatory descriptive terminology). The classes are listed in Table 1 and the resulting two-dimensional class-group taxonomy is illustrated in Fig. 3. I will comment on it in the order chondrites, irons, achondrites and then discuss more specifically the chondrite/achondrite transition.

## Chondrites

Although some of the Weisberg et al. (2006) chondrite "classes"—carbonaceous, ordinary, enstatite chondrites—may conceivably be petrographically defined, one would be at a loss at finding common petrographic traits for CCs, with the eponymous carbon (enriched only in CI and CM, essentially) having long ceased to be one (Krot et al., 2014). Rather, CCs can be defined by their isotopic composition lying in the C superclan (with e.g., $^{50}Ti$, $^{54}Cr$ enrichments; Kleine et al., 2020) and their chondritic petrography. That is, they are a group in our parlance. Within the C superclan, a "CR clan" has long been singled out (e.g., Weisberg et al., 1995) to encompass the CR, CH, and CB chemical groups, to which the IIC irons may be added (Budde et al., 2018). This clan (as hereby extended to nonchondrites) could be renamed the "CRHB group" (or "clan") to avoid notational confusion with the eponymous CR chemical group.[8] This would also allow us to compactly show the affiliation thereto of (non-CR) anomalous meteorites: For example, LEW 85332 (Torrano et al., 2021) could be reclassified as "CRHB3-an." Likewise, the "CM-CO clan" (hitherto only associated with bona fide CM and CO chondrites in the MetBullDB) and "CV-CK clans" (likewise, only CV and CK in MetBullDB) could be noted "CMO" and "CVK," respectively, for the same ends (hereby also avoiding interference with the lithology mixture conventions). Then, NWA 5958 could be ranked as CMO2-an (Jacquet et al., 2016; Torrano et al., 2021) and Ningqiang (Wang & Hsu, 2009) as CVK3-an. At present though, it is unclear whether the CMO and CVK clans correspond to distinct regions in isotopic space (in particular Cr and Ti,

---

[8]Such a concatenation of the suffixes of group names after the shared prefix merely generalizes the practice for irons (e.g., IIA and IIB combined as IIAB; Scott & Wasson, 1975), without any assumption on the number of parent bodies.

Table 1. Proposed classes and possible groups of member meteorites (not "groups in the class," because groups may straddle several classes).

| Class | Possible groups for meteorites in the class | Examples of anomalous meteorites in the class |
|---|---|---|
| RI-rich chondrites | CK, CV(red,ox), CO, CM, CL | NWA 5958 (C2-an; Jacquet et al., 2016) |
| Matrix-rich chondrites | R, K, CI, CR, | GRV 020043 (AL4-an; Li et al., 2018), NWA 8785 (EL3-an; Rindlisbacher et al., 2021) |
| Metal-rich chondrites | CH, CB(a,b), H [Portales Valley] | NWA 12273 (O3-an), NWA 5492 (NC3-an; Weisberg et al., 2015) |
| Chondrule-rich chondrites | H, L, LL, EH, EL | El Médano 301 (O4-an; Pourkhorsandi et al., 2017), LEW 87223 (E3-an) |
| Metachondrites | AL(ACA, LOD), WIN | Itqiy (EH8-an) |
| Irons (H, O[gg,g,m,f,ff, pl], D) (., silicate-, sulfide-rich) | IAB(−MG, sLL, sLM, sLH, sHL, sHH), IC, IIAB, IIC, IID, IIE, IIF, IIG, IIIAB, IIIE, IIIF, IVA, IVB | South Byron (C-an; Hilton et al., 2019), NWA 468 (AL-an; Li et al., 2018) |
| Pallasites | PMG, PES | Milton (C-an; Hilton et al., 2019) |
| Dunitic achondrites | BRA, DIO [olivine], ANG [NWA 8535], CHA | NWA 12319 (NCD-an) |
| Wehrlitic achondrites | BRA, URE ["olivine-pigeonite"], NAK | |
| lherzolitic achondrites | URE ["olivine-augite"], SHE ["poikilitic"] | NWA 13351 (O-an) |
| Harzburgitic achondrites | DIO [olivine], URE ["olivine-orthopyroxene"] | Dhofar 778 (NCD-an; Greenwood et al., 2017) |
| Orthopyroxenitic achondrites (., olivine-rich) | DIO, AUB, Mars [ALH 84001] | Shallowater (E-an; Zhu, Moynier, Schiller, Becker et al., 2021) |
| Noritic achondrites | DIO ["Yamato type B"], AUB [Bishopville], LUN [norite] | Al Bir Lahlou 001 (ung) |
| Gabbroic achondrites (., olivine-rich) | EUC [cumulate, Mg-rich], SHE ["gabbroic"], LUN [gabbro], ANG ["plutonic"] | NWA 7325 (NCD-an; Goodrich et al., 2017) |
| Basaltic achondrites (., olivine-rich) | EUC, ANG ["quenched"], SHE ["basaltic"], LUN [basalt] | Ibitira (NCD-an; Sanborn & Yin, 2014), NWA 011 (C-an; Warren, 2011a), NWA 8159 (Mars-an) |
| Troctolitic achondrites | LUN [troctolite] | |
| Anorthositic achondrites | LUN [anorthosite] | |
| Polymict achondrites | HOW, MES(−A,B,C), LUN ["mingled"], Mars [NWA 7034] | Chaunskij (ung; Petaev et al., 1992) |

In the first two columns, parentheses indicate possible subdivisions separated by commas (a "." meaning no qualifier, petrographic types of chondrites and mesosiderites being omitted for clarity). Brackets in the second column may specify what specific variant (or single member) of the group may be comprised in the class (between quotes when not a MetBullDB expression) and which are superseded by the class information. Example anomalous meteorites in the class are listed in the third column, with (group-wise) classifications normally based on Meteoritical Bulletin Database (2021), unless specific references (e.g., for isotopically defined high-level groups/"(super)clans") are indicated. ACA = acapulcoite, AL= Acapulco-Lodran group, ANG = angrite, AUB = aubrite, BRA = brachinite, CHA = chassignite, DIO = diogenite, HOW = howardite, LOD = lodranite, LUN = Lunar, MES = mesosiderite, NAK = nakhlite, PES = pallasite Eagle Station group, PMG = pallasite main group, RI-rich = refractory inclusion-rich, URE = ureilite, WIN = winonaite. For iron classes: H = hexahedrite, O = octahedrite (coarsest = Ogg, coarse = Og, medium = Om, fine = Of, finest = Off, plessitic = Opl), D = ataxite.

which, unlike oxygen, are unaffected by aqueous alteration; Schrader & Davidson, 2017). For example, Zhu, Moynier, Schiller, Alexander, et al. (2021) report indistinguishable chromium-54 excesses in CM, CO, and CV chondrites (with CK being systematically lower). Torrano et al.'s (2021) data set CM at higher chromium-54 excesses than CO, but the latter are not isotopically closer to CM than (literature) CV and CK. Schrader and Davidson (2017) argued that the chondrule populations incorporated in CM and CO chondrites were not the same (i.e., these chondrites did not merely differ by the abundance of matrix upon accretion). So the genetic value of the CMO and CVK clans is somewhat uncertain and their definition remains at any rate dependent on petrographic criteria (e.g., chondrule size, CAI abundance). They are outlined in dotted lines in Fig. 3.

As to noncarbonaceous chondrites (NCCs), ordinary chondrites (OC) are best viewed as the chondritic subset of an isotopically definable O group (or "clan"), which also contains IIE and IVA irons (Kleine et al., 2020; Weisberg et al., 2006). Same for EC with respect to the E group (which also comprises aubrites) or NCCs as a whole among NCs.[9] The

[9]So technically the additional "C" in the foregoing group acronyms has a differential (more restricted) meaning that should be attended to (as in Fig. 3).

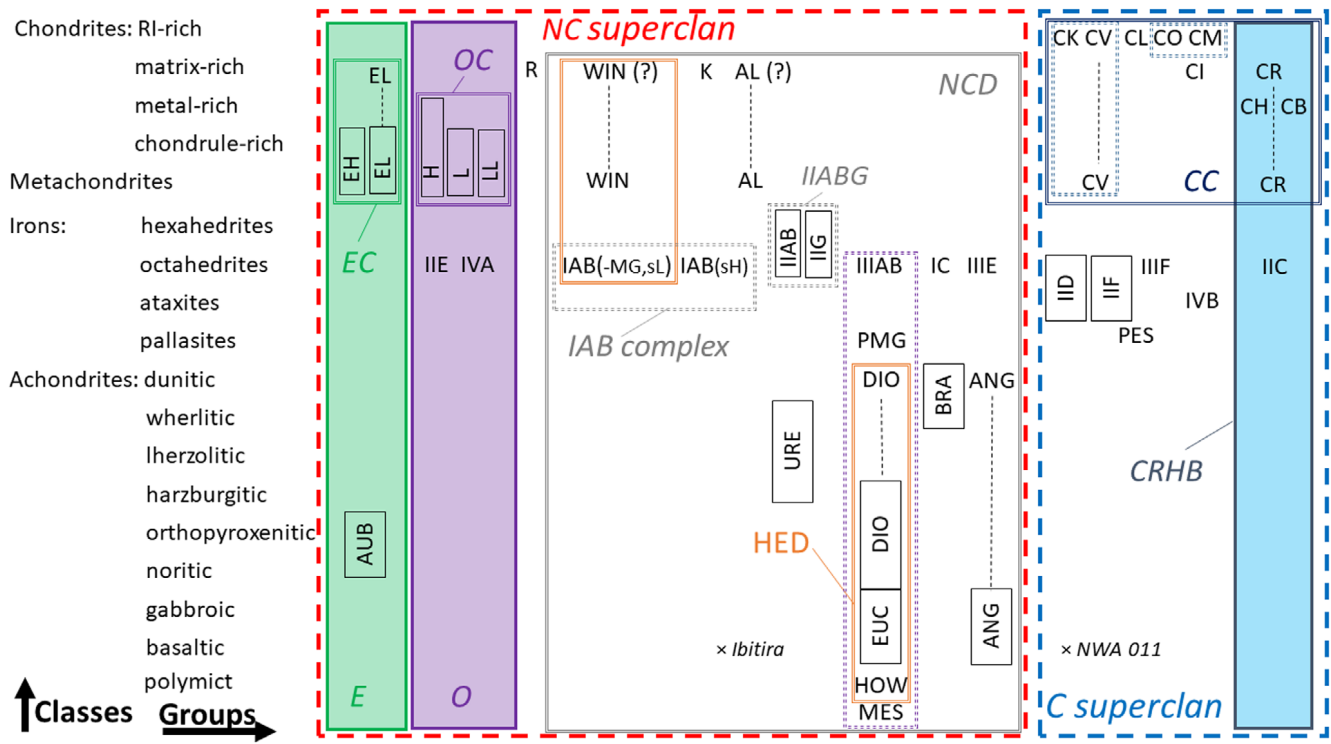

Fig. 3. A petrology-genetics diagram for current asteroidal meteorite groups with known C/NC affinities (e.g., Kleine et al., 2020; Weisberg et al., 2006). The abscissa is arbitrary and proximity between two groups does not imply genetic relationships unless they are explicitly boxed in a common higher level group (boxed low-level groups, whose names are written in the vertical direction, are meant to span the whole class range encompassed; gaps are bridged by a dotted line). The hypothetical assignations of chondritic winonaites and acapulcoites to the matrix-rich chondrite class are discussed in the section Chondrites. Here, the ordinary chondrite (OC) clan is distinguished from the more inclusive O group, which comprises differentiated meteorites (e.g., IIE irons); same for EC versus E and CC versus C (see the Chondrites section). I show in dotted magenta the nameless Rubin and Ma (2021) HED-MES-PMG-IIIAB "superclass," which is a group in our parlance.[10] All mentioned high-level groups are meaningful whether they represent one or several primary parent bodies, hence the impossibility to define a real hierarchy among them except locally for nested groups. While only established groups are exhaustively shown, any anomalous meteorite must land somewhere in this diagram, in its class, and a more or less broadly defined group (e.g., C, NC, O, E) and thus lend itself to a minimum standardized taxonomic characterization. As an illustration, the "eucrite-type" Ibitira and NWA 011 meteorites have been plotted in the same basaltic achondrite class as bona fide eucrites but they are discriminated from them on the group axis as "NC-an" and "C-an," respectively. Abbreviations are the same as in Table 1. (Color figure can be viewed at wileyonlinelibrary.com.)

boundaries of the E and O groups are a matter of convention, but in the MetBullDB, the O group has already been extended to meteorites more reduced (and [16]O-richer) than H chondrites (e.g., NWA 13411 mentioned in the previous section), so it might be agreed to stop only wherever the chondrite mineralogy changes qualitatively to an EC one (e.g., with non-iron sulfides such as oldhamite, etc.). R chondrites, whose Cr isotopes resemble ECs but which are [16]O-poorer than even OCs, are best left out of either clan, but more isotopic measurements (e.g., Ti) would be worthwhile. A few NC chondrites, such as K chondrites (indeed NCCs judging from [50]Ti deficits measured by Niemeyer [1985, 1988] and [48]Ca deficits reported by Prombo and Lugmair [1987] or GRV 020043 [Li et al., 2018] show negative $\Delta^{17}$O). That part of the NC isotopic space is actually far from anecdotal, for it is occupied by most *differentiated* meteorites (e.g., Kleine et al., 2020), so it is worth a proper name. I propose "NCD," standing for "Negative Chromium-54 excesses and Delta oxygen-17," or, indifferently "normal NC Differentiated meteorite signature," as shown in Fig. 3. Few planetesimals

---

[10]This follows generally from Rubin and Ma's [2021] definition of a superclass as a set of groups "formed on different asteroids, but presumably in the same general locale of the solar system." Their other "superclass," that of the enstatite meteorites, corresponds to the E group in this paper. It may be noted that "classes" in the sense of Rubin and Ma [2021] also carry a genetic connotation, as grouping "one or more clans that formed in the same general region of the solar system." It is seen that there is little qualitative difference between the definitions of these two levels of the hierarchical system of Rubin and Ma [2021], and one of my chief points here is that it cannot be otherwise.

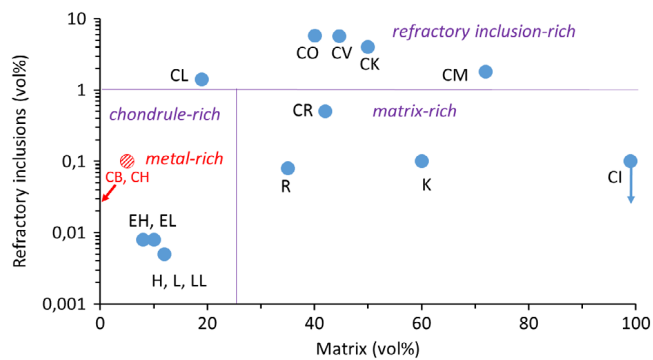

Fig. 4. Matrix versus refractory inclusion (calcium-aluminum-rich inclusions and amoeboid olivine aggregates) abundances in chondrite groups. The different chondrite classes can be discriminated on this diagram, except metal-rich chondrites (hatched circles) which are distinguished on a third axis (metal abundance) that may envisioned to be perpendicular to the plane of the figure. CO and CV data from Ebel et al. (2016), CM (Murchison) data from Fendrich and Ebel (2021), CL data from Metzler et al. (2021), refractory inclusion abundances from Dunham et al. (2021) for OCs and ECs, and Rout and Bischoff (2008) for R chondrites, and remaining matrix data in priority from Rubin (2010) and then Krot et al. (2014) for the rest. For K chondrites, the 0.1 vol% refractory inclusion abundance is technically an upper bound from Weisberg et al. (1996), but the >10 CAIs (<50–400 μm across) found in the two Kakangari thin sections of Prinz et al. (1989) and the 0.35 vol% found by Barosch et al. (2020) in theirs suggest it is not far from the truth. (Color figure can be viewed at wileyonlinelibrary.com.)

accreted in the NCD space–time region thus apparently escaped differentiation.

With our traditional high-level chondrite taxa and some newer ones cast as groups, classes can be devised afresh to evaluate the efficiency of the production (and transport) of high-temperature components (nebular condensates and chondrules) in them. Four classes may be proposed as follows:

*Metal-rich chondrites* contain more than 15 vol% metal.
*Refractory inclusion-rich (or RI-rich) chondrites* have more than 1 vol% refractory inclusions.
*Matrix-rich chondrites* have more than 25 vol% matrix.[11]
The other chondrites are *chondrule-rich chondrites*.

More precisely, should a chondrite satisfy several of the above criteria, the earliest in the list shall have priority (so an RI-rich chondrite remains an RI-rich chondrite irrespective of matrix abundance, e.g.). The chondrite chemical groups (or rather their averages) are distributed among the classes in Fig. 4. It is remembered that classes refer to the primary (i.e., type 3) rock, whose petrography may be more or less hypothetically reconstructed. For example, metamorphosed CK may have lost recognizable matrix or refractory inclusions (Chaumard et al., 2014) and their assignation to refractory inclusion-rich chondrites relies on their supposed link with unequilibrated members of their group. When such an identification is not possible, for example, for an ungrouped type 6 sample, the assigned class can only be a broad "chondrite" (but would be redundant with the petrological type). Let us now motivate more specifically the classes defined above.

The metal-rich chondrites, which comprise, for example, CH and CB chondrites, are set apart in priority because they are the "least chondritic" of chondrites. The 15 vol% metal threshold basically ensures supersolar Fe/Si ratios, which is unlike the solar or subsolar ones shown by the vast majority of chondrites (e.g., Hutchison, 2004). Now, CH chondrites still show nonvolatile major element (Ca, Al, Mg, Si, Fe, Ni) ratios within a factor of 2 of solar (CI) composition. This however breaks down for CBs unless one restricts attention to ratios between elements of like geochemical affinities (lithophile or siderophile). Rather, CBs remain chondrites inasmuch as they contain chondrules (silicated and metallic), even though those chondrules, as well as part of those in CH, may have different formation scenarios than their "mainstream" counterparts, possibly in an impact plume (Krot et al., 2005)—but a descriptive taxonomy should not judge.

The refractory inclusion-rich chondrites are singled out next because they fossilize the first generation of solar system solids (either intact or as relics or bulk compositional signatures in chondrules, e.g., Jacquet, 2021; Marrocchi et al., 2019). The threshold chosen (1 vol% refractory inclusion) ensures that refractory inclusions appreciably contribute (by more than about a tenth) to the refractory element budget (given that they are enriched in them by more than one order of magnitude relative to CI; e.g., Grossman, 2010) and lies in a neat hiatus among CCs: in fact, RI-rich chondrites comprise all CCs except the CRHB clan and CIs.

The matrix-rich chondrites seem overall less depleted in RI than chondrule-rich chondrites, even in the NC superclan. The few known NCD chondrites seem to be all matrix-rich; such are the K chondrites and GRV 020043 (38 vol% matrix), which may be an acapulcoite precursor (Li et al., 2018); also, the winonaite NWA 725, which really is a type 5 chondrite, only contains 5 vol% relict chondrules (Zeng et al., 2019) in contradistinction to H5 and H6 chondrites which contain 22 and 11 vol%, respectively (Rubin et al., 2001). Perhaps NC matrix-rich chondrites accreted earlier than their chondrule-rich counterparts and were thus more prone to at least partial differentiation because of more abundant aluminum-26. Chondrules may have been produced mainly in the interim, and refractory inclusions, already rarer than in CCs, would have been further depleted (e.g., by inward drift; e.g., Pignatale et al., 2018) to the very low abundances observed in (chondrule-rich) ordinary and ECs (e.g., Dunham et al., 2021) (Fig. 4). At any rate, matrix-rich chondrites are representative of regions of poor chondrule formation and/or transport efficiencies which are yet to be put in astrophysical context.

---

[11]Matrix is somewhat difficult to define, but so are (correspondingly) chondrules, in fact, as coarse isolated silicate or metal grains may also count as chondrules (Jacquet et al., 2021; Zanda et al., 1994). The boundary between the two should be some critical grain size (either absolute or as a fraction of a typical chondrule size).

The chondrite classes are listed in Table 1 and Fig. 3 roughly in order of decreasing Libourel et al. (2017) "primitiveness," that is, decreasing refractory inclusion and increasing chondrule abundances.

## Irons

For irons, I essentially copy-paste the structural class/chemical group paradigm (e.g., Buchwald, 1975). While, de facto, the MetBullDB only recognizes the "iron" class (in expressions such as "iron, IVB" or "iron, IIAB"), this appears disproportionately coarse when compared with stone meteorite taxonomy. I thus suggest reviving the older structural classes, at least at the highest level—hexahedrite, octahedrite, ataxite (see Fig. 3)—which incidentally, in being chiefly controlled by Ni content, roughly correlates with the redox state of the parent body, with more oxidized conditions apparently prevailing for C irons (e.g., Hilton et al., 2022; Spitzer et al., 2021). For many irons, structural classes can be extracted from Buchwald's (1975) compilation. It is remembered that Buchwald (1975) slightly adjusted the boundaries of the structural classes so that chemical groups of the time could fit entirely in only one, but now class transversality is the rule (e.g., IIAB irons can be hexahedrites or [coarsest] octahedrites; Scott, 2020) (Fig. 3).

The distinction between fractionally crystallized and nonfractionally crystallized irons is too theoretical for a descriptive class dimension, but is nonetheless useful. Perhaps, these irons could be respectively called "fracrystallized" and "non-fracrystallized" to rival in brevity with the older, now improper expressions "magmatic" and "nonmagmatic" still in wide use, despite both varieties of irons being igneous (Scott, 2020). Although ignored as such in the official classification, non-fracrystallized irons not uncommonly abound in silicate (e.g., Ruzicka, 2014) or troilite inclusions (frequent in the non-fracrystallized irons; e.g., Ruzicka, 2014) which may be added to the class descriptors (e.g., "iron, silicated" or

"octahedrite, sulfide-rich"), if either exceeds 10 vol%. The silicated irons have also prompted me to list pallasites, which are but a variant thereof, alongside irons (e.g., Jacquet, 2001). In fact, evidence that pallasites formed from the mixing of metal with preexisting dunites (Scott, 1977) could recast them as "dunite-irons" (following the Lithology Mixing section), or "harzburgite-irons" for the most pyroxene-rich ones (e.g., Rubin & Ma, 2021) but Table 1 remains terminologically conservative.

Many non-fracrystallized irons are affiliated to the "IAB complex" (Wasson & Kallemeyn, 2002). Although its component "main group" and five "subgroups" are certainly all groups in our parlance, the taxonomic status of the ensemble (which also comprises many ungrouped irons) is open to question in the binominal scheme. The Wasson and Kallemeyn (2002) *chemical* definition (essentially, irons not too depleted in moderately volatile siderophile elements) would be closer to a class, as it would not guarantee a genetic significance more than the volatile depletion-based definitions of the older iron "groups" II, III, and IV (e.g., Lovering et al., 1957; Scott & Wasson, 1975), which straddled both NC and C superclans (e.g., Kleine et al., 2020). That definition is, however, not consistently implemented in that the earlier established (also non-fracrystallized) IIE iron group, which mostly satisfy their criteria (or only narrowly miss some; data of Wasson, 2017), is not included in the IAB complex. Now IIE irons belong to the O clan whereas the IAB complex irons subjected to isotopic analysis thus far seem to lie all in the NCD group introduced in the previous subsection (e.g., Dey et al., 2019). Perhaps then, the "IAB complex" may be redefined as the set of those irons satisfying the Wasson and Kallemeyn (2002) chemical criteria belonging to the NCD group, or some more restricted subset thereof (e.g., Wasson and Kallemeyn 2002 "classic range" $-0.68‰ \leq \Delta^{17}O \leq -0.3‰$). So the IAB complex may indeed be a group in our parlance (as outlined in dotted line in Fig. 3), provided isotopic outliers are excluded.

Among fracrystallized irons, the association between IIAB and IIG irons proposed by Wasson and Choe (2009) is outlined in dotted line in Fig. 3 and may be referred to as IIABG.

## Achondrites

Since the Rose–Tschermak–Brezina terms for achondrites have been understood here to refer to groups, class names must be different. Indeed, expressions such as "brachinite-like" (Day et al., 2012) would be too ambiguous (and anyway are generally meant to *exclude* the eponymous groups; although the "eucrite-type" achondrites of Mittlefehldt et al. [2022] do include bona fide eucrites).

One could think of resurrecting Prior (1920)-style mineral designations (e.g., hypersthene achondrites, plagioclase–pigeonite achondrites, etc.), which are echoed in modern designations of some variants such as "diogenite–olivine" or "olivine–pigeonite," "olivine–augite," and "olivine–orthopyroxene ureilites" (though the latter three do not appear in MetBullDB; see Goodrich et al., 2004). A more quantitative solution though is to resort to rock names from IUGS (International Union of Geological Sciences) terminology (e.g., Le Bas & Streckeisen, 1991), as already consistently done for lunar meteorites (see also Le Bas, 2001), and proposed for some asteroidal achondrite groups (e.g., diogenites; Beck & McSween, 2010).

Use of terrestrial terminology should not be construed to imply petrogeneses for achondrites and their terrestrial counterparts, for example, as to the role of plate tectonics, etc. (Day et al., 2009). In fact, IUGS terminology itself was meant to be essentially descriptive, in terms of modal mineralogy and/or bulk chemistry (Le Bas & Streckeisen, 1991). Still, there is a distinction between volcanic and plutonic rocks (with the ambiguous "hypabyssal" or "subvolcanic" rocks) which may be no less equivocal for some achondrites than for terrestrial rocks, for example, eucrites whose thermal metamorphism may be due to reheating after igneous crystallization rather than formation at depth (e.g., Mittlefehldt, 2015) or the "poikilitic shergottites" of Walton et al. (2012). For eucrites, the line is here drawn between their noncumulate ("basaltic") and cumulate varieties (Table 1).

In detail, cosmochemical practice may somewhat deviate from strict IUGS terminological boundaries. For example, the name "basalt" liberally given to many extrusive achondrites does not always agree with terrestrial standards (e.g., Keil, 2012; Le Bas, 2001). In due course, stricter adherence to IUGS terminology may be the way to go, but it would also be conceivable to adapt terminological boundaries to natural conditions on asteroids rather than on Earth. Here, as a first step, I have elected to use IUGS rock names only adjectively in phrases like "basaltic achondrite" (instead of simply "basalt") so as to signal some leeway as well as their extraterrestrial nature. When the IUGS term is a compound expression with a mineral name, for example, "olivine basalt," its meteoritic counterpart will be of the form "olivine-rich basaltic achondrite." In practice though, when the (extraterrestrial) context is clear, direct use of IUGS substantives (e.g., "asteroidal basalt," "HED olivine orthopyroxenite," etc.) should be unobjectionable.

Among achondrite group names, "chassignites" may be considered essentially as an historical accident of meteorite recovery. Indeed, had Chassigny not fallen in 1815, the two other chassignites NWA 2737 and NWA 8694 (not meeting the five-member threshold introduced meanwhile for establishing a new group anyway) would have been directly lumped with the genetically related nakhlites (while remaining distinguishable classwise as dunitic achondrites) in an epoch where "shergottites" were accreting a petrographically diverse suite of Martian mafic rocks. I shall keep my promise to conserve existing group names in this section (in particular Table 1), but will return to the topic of further nomenclature simplification in the conclusion.

While a mesosiderite can be ranked classwise as basalt–orthopyroxenite–iron (or orthopyroxenite–iron if of type C), "mesosiderite" is here considered to be a *group* name, since present-day mesosiderites[12] exhibit a consistent (HED-like) isotopic signature (Greenwood et al., 2017). Still, some so-called "anomalous mesosiderites" (e.g., the isotopically distinct Chaunskij; Petaev et al., 1992) would have yet to be excluded from the taxon.

The achondrite classes are listed in Table 1 and Fig. 3. The order very roughly evokes the stratigraphy of a differentiated parent body (~irons, Ca-poor achondrites, Ca-rich achondrites, and breccias). Of course, this should not be taken too seriously and does not claim more than mnemonic value.

### The Chondrite/Achondrite Transition

Primitive achondrites, which may be defined as residues of (silicate) partial melting of chondritic protoliths (e.g., Tomkins et al., 2020), are not included *as a class* in the taxonomy. This is because the restite nature of some achondrites may be contentious (e.g., some brachinites and ureilites may be cumulates; Day et al., 2012; Goodrich et al., 2004, 2009; Keil, 2014), and as such unsuitable for a descriptive taxon (similar to the fraccrystallized/non-fraccrystallized iron distinction). Many "evolved" primitive achondrites, such as most brachinites and ureilites, lend themselves to the taxonomy of ultramafic plutonic rocks (e.g., dunitic achondrites, wehrlitic achondrites, etc.) discussed in the previous subsection. Yet opaque richer (say with >5 vol% metal+sulfide[13]), achondritic textured meteorites do not, and for this class, I adopt the term "metachondrite" originally coined by Irving et al. (2005).[14] Metachondrites as descriptively defined here include not only primitive achondrites such as acapulcoites–lodranites and winonaites but also chondritic impact melts, which may be occasionally difficult to unequivocally identify as such (e.g., Rubin, 2007). So metachondrites include products of static *and* shock metamorphism, both consistent with the etymology, which incidentally leaves suitably ambiguous whether the meteorite actually counts as a chondrite or an achondrite.

---

[12]After the exclusion of many historical outliers, such as Weatherford, Bencubbin (both CBs), Udei Station (IAB-an), Winona (WIN) which Mason (1962) was still listing among mesosiderites.

[13]Judging from analyses by Jarosewich (1990), for residues from oxidized, iron-depleted chondritic precursors (e.g., LL), it may be necessary to lower this threshold, but only on condition of >5 vol% plagioclase (i.e., limited partial melt extraction), to avoid including (possibly cumulate) brachinites (e.g., Keil, 2014) or ureilites. So all demonstrably type 7 meteorites (as redefined later) would be metachondrites. For those, a threshold of 3 or 4 vol% may be sufficient to distinguish from evolved achondrites which Jarosewich's (1990) analyses suggest have generally sufficiently separated from metal and sulfide (barring impact mixing).

Of course, more information is not hereby barred from entering the classification. If the metachondrite is recognized as a chondritic impact melt of a chondrite, this may be embodied in a C-S7 shock stage (Stöffler et al., 2018), and the metachondrite classwise identification becomes superfluous. If, instead, it is identified as the result of progressive heating (regardless of whether part of the heat is due to a nearby impact; Tait et al., 2014; Tomkins et al., 2020), that information may be encoded in the petrologic type, going beyond the type 6 of Van Schmus and Wood (1967).

Indeed, I adopt here the type 7 definition of Tait et al. (2014), which corresponds to the onset of Tomkins et al. (2020)'s "Sub-calcic Augite Facies," and indicates incipient partial melting of silicates. This is evidenced by, for example, interconnected plagioclase ± pyroxene networks and virtual lack of relict chondrules (Tait et al. (2014) tentatively suggest <1 cm$^{-2}$ in their study of the Watson 012 H chondrite; while Zhang et al. (1995) had proposed <5 cm$^{-2}$ for ECs). This type 7 includes most acapulcoites and winonaites (Zeng et al., 2019). Some chondrites previously classified as type 7 that are merely extensions of type 6 (with relict chondrules; e.g., Irving et al., 2019) may be lumped back therewith or be given a transitional type 6/7.

A *type 8* is introduced here to denote significant extraction of silicate melt,[15] as a generalization of the classical distinction between lodranites and acapulcoites. One may set a quantitative threshold at 5 vol% plagioclase (judging from modal data of the Acapulco–Lodran group compiled by Keil & McCoy, 2018). Aside from lodranites, the type 8 thus defined includes, for example, Itqiy (which appears as EH7-an in the MetBullDB), most brachinites (except the near-chondritic Brachina which should be type 7; Keil, 2014) and ureilites. Other primitive achondrites showing evidence for significant silicate melt mobilization but not meeting the above threshold, such as transitional acapulcoites/lodranites or "enriched" acapulcoites (Keil

Table 2. Possible replacements of some traditional taxonomic terms.

| Traditional taxon name | Binominal equivalent |
| --- | --- |
| Acapulcoite | AL7 or AL6 (i.e., type 7 or 6 Acapulco–Lodran [meta]chondrite) |
| Chassignite | Dunitic nakhlite |
| Diogenite | Vestan orthopyroxenite (or dunite, or norite) |
| Eucrite | Vestan basalt (or gabbro) |
| Howardite | Vestan basalt–orthopyroxenite breccia (or "diogenite–eucrite") |
| Lodranite | AL8 |
| Mesosiderite (as a class) | (basalt-)orthopyroxenite–iron |
| Pallasite (as a class) | Dunite–iron (or harzburgite–iron) |

The replacement of the acapulcoite/lodranite distinction as petrologic type is discussed in the Petrologic Types subsection. The name "nakhlite" is hereby proposed to extend to the genetically related chassignites (and thus to keep "wehrlitic achondrites" only as a *default* class). If the "Vestan" adjective given to howardites–eucrites–diogenites seems to suggest too strongly their arguably conjectural parent body, it might be replaced by "HED" (which would be only *historically* analyzable as a taxon name acronym). For pallasites, the substitution is only meant classwise, for "pallasite" could still linger in the names of the associated groups ("main group pallasites," "Eagle Station pallasites"), which ought to be changed only in case of unification with existing groups. The name "mesosiderite," already rejected *as a class* in Table 1 (see the Achondrites section), is likewise spared for the foreseeable future *as a group designation* (unless, again, it can be someday superseded by an hyphenated form of the sort "HED-IIIAB" if the component lithologies can be identified to preexisting groups).

& McCoy, 2018) or the Sahara 02029 and Tierra Blanca winonaites (Zeng et al., 2019) may be ranked as type 7/8. Watson 012, Tait et al.'s (2014) prototype for their type 7 redefinition, shows net gain of basaltic melt relative to H6 chondrites and might likewise be ranked as H7/8.

Such an extension of Van Schmus and Wood's (1967) scale to partial melting should not appear illegitimate outright since that scale already concatenated two different processes—namely, aqueous alteration (1–2) and thermal metamorphism (3–6). In fact, partial melting may already occur at type 6 for metal–sulfide (Tomkins et al., 2020). The point of this extension is to allow a consistent subclassification of those groups which straddle chondrites and achondrites, with no artificial nomenclatural discontinuity. Thus, while many winonaites are WIN7, some have type 5 and 6 chondritic textures (Zeng et al., 2019) and should be referred to as WIN5 and WIN6, respectively. Likewise, while most acapulcoites and lodranites can be recast as AL7 and AL8, respectively, the chondrule-bearing acapulcoite GRA 98028 (e.g., Schrader et al., 2017) would be an AL6, and GRV 020043 (Li et al., 2018) would be an AL4. Conversely, there is no objection in considering a LL7 as a primitive achondrite even though the expression "LL7 chondrite" may be further tolerated (the word "metachondrite," as said, circumvents this semantic issue).

---

[14]If with a more restricted meaning, but Irving et al. (2019) have anyway withdrawn it.
[15]This is distinct from the type 8 proposed by Irving et al. (2019), which essentially corresponds to the type 7 herein.

## PERSPECTIVES

However rationalized the binominal taxonomy is argued to be, how realistic is it to expect that the meteoritical community adopt it? There would be, in fact, little to adopt in the short term. In the particular implementation sketched in the foregoing section, the groups would be strictly conserved (notwithstanding the exclusion of isotopic outliers), with classes merely superimposing a standardized petrological grid on them. Nothing here would make past literature unintelligible to future meteoriticists hypothetically trained in the binominal formalism, or vice versa. In fact, no suggestion is entertained to legislate on scientific writing so long it is analyzable in NomCom-approved class/group terms. Classes would only have to be systematically distinguished in the MetBullDB, and this only whenever they differ from the default expected for a given group. So absent any discontinuity in the scientific literature, general acceptability might not be much of an issue.

Yet, the advantages of the binominal scheme would not fail to be noticed in due course:

A better characterization of lithologically diverse groups (especially among differentiated meteorites),
A minimum standardized taxonomy of ungrouped meteorites,
The availability of their genetic affinities, and
A clarification of the "genetic" language.

So in the short term, the advantage of the binominal scheme chiefly lies in the characterization of anomalous samples, and its handling of established groups may be merely viewed as a proof of concept of its consistency. Yet the latter may have pedagogical value. A decidedly class-oriented introduction to meteorites may indeed prove advantageous for students, by first laying stress on petrological processes (pre- and post-accretion) rather than the more contingent genetic associations sampled in the meteoritical record. In fact, the latter associations may be quite compactly presented in "petrography-genetics" diagrams such as Fig. 3 (or variants replacing its arbitrary $x$ axis by some isotopic ratio), once isotopic biplots have made the C/NC dichotomy familiar (Kleine et al., 2020; Warren, 2011a, 2011b). The various affinities between different low-level groups discussed in the literature may be seen at a glance. Figure 3 also illustrates the NC(D) affinities of most differentiated meteorites (Warren, 2011a, 2011b), and among chondrites themselves, the less primitive (more chondrule-rich, refractory-poor) textures of the NC superclan. The "dunite problem," that is, the overrepresentation of iron cores compared to mantles (and crusts) in the meteoritical record (e.g., Scott et al., 2015), is also obvious. Needless to say, all these are fundamental observations that cannot be given too much visibility.

On a longer term, once classes prove sufficiently adapted to the petrographic diversity of meteorites, named groups may evolve, case by case, toward more purely reservoir-related (~isotopic) concepts, with greater class transversality, from the already familiar NC and C superclans down to single parent body-wide scale.[16] For example, IAB-MG (or sL) irons could become "iron winonaites" (e.g., Worsham et al., 2017). "Ungrouped" meteorites associated isotopically and petrogenetically with established groups may be lumped with them—but in their appropriate classes, for example, GRA 06128 and GRA 06129 (e.g., Day et al., 2012) could become "andesitic brachinites." The touchstone for such group (or grouplet) mergers would not be so much whether the preexisting groups derive from the same primary parent body—since, again, this cannot be ascertained for the preexisting groups individually—but whether the parent bodies of the preexisting groups are indeed *practically* indistinguishable, isotopically and petrogenetically.

If this is the case, at least one of the previous group names becomes unnecessary. Table 2 lists taxon names which *in principle* could be already eliminated, although the previous section (e.g., Table 1) has stuck to existing groups, pending evolving practice of the community. Yet the elimination of the taxon names of Table 2— which are mere historical accidents of the chronology of meteorite falls and finds—would complete the nomenclature simplification undertaken by Prior (1920) and compensate somewhat for the emergence of new groups in the crowded memory of 21st-century meteoritics students. Now some of the old names could linger as synonyms for the relevant class–group intersections (or lithology mixtures). It is remembered that Van Schmus and Wood (1967) had upheld the older expressions "bronzite chondrites," "hypersthene chondrites," or "amphoterites," reinterpreted to designate equilibrated (i.e., type 4–6) H, L, and LL chondrites, respectively, even though these venerable terms eventually faded into oblivion. The binominal scheme would thus become more truly "two-dimensional" (petrography–reservoir) in the same sense the Van Schmus and Wood (1967) scheme is for the primary/secondary property duality among chondrites. Whatever the future holds in store, these possible evolutions bear witness to the flexibility of the binominal reconceptualization of meteorite classification.

---

[16]Or even below, e.g., for "enriched"/"intermediate"/"depleted" shergottites, which qualifiers are already treated independently of the textural classification (e.g., McSween & McLennan, 2014).

*Acknowledgments*—I am grateful to Alan Rubin and Jeffrey Grossman for fruitful taxonomic discussions (following an earlier text on the treatment of anomalous meteorites), emphasizing the textural/genetic mixture of the current classification (justifying a twofold anomaly concept), as well as a review of the former on this manuscript, which led to improvements in the achondrite taxonomy. A detailed review by Michael Weisberg greatly improved the taxonomy of chondrites and the overall rationalization of the system. I am also indebted to the excellent "Systematic Classification of Meteorites through Photographs" of David Weir (2021) as a useful guide to relevant scientific literature.

*Data Availability Statement*—No new data were produced for this work.

*Editorial Handling*—Dr. Michael Zolensky